# Pulsar Recoil and Gravitational Radiation due to Asymmetrical Stellar Collapse and Explosion

Adam Burrows and John Hayes

*Departments of Physics and Astronomy*

*University of Arizona*

*Tucson, Arizona, USA 85721*

## Abstract

New data imply that the average velocity of radio pulsars is large [1]. Under the assumption that these data imply that a pulsar is born with an "intrinsic" kick, we investigate whether such kicks can be a consequence of asymmetrical stellar collapse and explosion. We calculate the gravitational wave (GW) signature of such asymmetries due to anisotropic neutrino radiation and mass motions. We predict that any recoils imparted to the neutron star at birth will result in a gravitational wave strain, $h_{zz}^{TT}$, that does not go to zero with time. Hence, there may be "memory" [2] in the gravitational waveform from a protoneutron star that is correlated with its recoil and neutrino emissions.

Typeset using REVTEX

# I. INTRODUCTION

Recent data on pulsar proper motions [1], a recalibration of the pulsar distance scale [3], and a general recognition that previous pulsar surveys were biased towards low speeds [4] imply that many pulsars have high proper motions. Mean three-dimensional galactic speeds of 450±90 km s$^{-1}$ have been estimated [4], with measured transverse speeds of individual pulsars ranging from zero to as much as ∼1500 km s$^{-1}$. It has long been known that impulsive mass loss in a spherical supernova explosion that occurs in a binary can impart to the nascent neutron star a substantial proper motion that reflects its progenitor's orbital speed [5,6]. However, theoretical studies of binary evolution through to the supernova phase fail to reproduce velocity distributions with the requisite mean and dispersion [7,8].

Consistent evolutionary scenarios for the binary pulsars PSR B1534+12 and PSR B1913+16 [9], the observed misalignment of PSR B1913+16's spin axis [10], a model for Her X–1 [11], and the large eccentricities of Be/neutron star binaries [12] all buttress the conclusion that neutron stars are given an extra kick at birth. Recently, Caraveo [13] and Frail, Goss & Whiteoak [14] have identified pulsars with young supernova remnants (SNR's) in a majority of SNR's with putative ages less than 20,000 years, but only on the assumption that the pulsars have relatively high transverse velocities with an average value of ∼500 km s$^{-1}$.

Related to the emerging SNR/pulsar associations are the new data on the "systemic" velocities of young supernova remnants [15,16]. It is observed that the "center-of-mass" velocity of oxygen clumps in four of these explosions is different from that of the local ISM by on average more than 500 km s$^{-1}$ [16]. These data imply that the explosions themselves were asymmetric, but inhomogeneities in the ISM and a high progenitor speed can not yet be ruled out. It would be difficult to explain the jagged optical and IR line profiles of SN1987A, the intrinsic polarization of spectral features in SN1993J and SN1987A, and the oblateness of recent HST images of SN1987A without recourse to asymmetries during the explosion itself. In addition, Utrobin *et al.* [17] interpret the "Bochum" event in H$\alpha$ in SN1987A with a $10^{-3}$ $M_\odot$ shard of $^{56}$Ni moving at $\sim$4700 km s$^{-1}$. Explosion asymmetries and the systemic SNR speeds may be (anti)correlated with the recoil of a nascent neutron star.

## II. RECOILS IN THE CONTEXT OF ASYMMETRIC COLLAPSE AND EXPLOSION

In the past, an off-center (and rotating) magnetic dipole [18] and anisotropic neutrino radiation [19,21,22] have been invoked to accelerate neutron stars. It is the contention of this paper that asymmetries in the collapse and before and after the reignition of the supernova have the potential to impart to the core large recoils. That core-collapse supernovae are subject to Rayleigh-Taylor instabilities has been independently and convincingly demonstrated by at least three theoretical groups [23–25]. The pre-explosion core experiences a convective "boiling" phase behind the temporarily stalled shock, and the explosion, when it occurs, erupts in bubbles, fingers, and plumes. To date, these radiation/hydrodynamic calculations have been performed in only two dimensions (with axial symmetry) and with a variety of simplifying approximations. Nevertheless, this class of 2-D simulations can help theorists explore the potential role of hydrodynamics and neutrinos in imparting proper motions. The 2-D calculation of Burrows, Hayes, and Fryxell [23] was done assuming that the collapsing Chandrasekhar core and implosion were *spherical*. Even in this case, the core received stochastic impulses sufficient to shake and rotate the residue.

Interestingly, recent hydrodynamic calculations [26] of convection during shell oxygen and silicon burning in massive stars and recent theoretical arguments [27] suggest that just before collapse *asymmetries* in density, velocity, and composition can be large. Furthermore, rotation might interact with convection to further distort the core [28]. The upshot might be an asymmetrical, aspherical collapse. If the amplitudes of the asymmetries in density or velocity are not negligible and if a significant low-order mode ($\ell = 1$) exists at the onset of collapse, the young neutron star can receive a large impulse during the explosion in which it is born.

To investigate this hypothesis, we conducted an exploratory and experimental calculation of aspherical collapse, using the code described in Burrows, Hayes, & Fryxell [23]. In this simulation, we artificially decreased by 15% the density of the Chandrasekhar core exterior to 0.9 $M_\odot$ and within 20° of the pole. Such a perturbation in the 15 $M_\odot$ progenitor that we used [29] amounts to a core-wide mass dipole anisotropy of less than 0.1% (and on any given mass shell of less than 2%). The initial total quadrupole anisotropy is 0.008 (normalized to 2/3). This calculation was done in 2-D with azimuthal symmetry, but $\theta$ ranged between 0° and 180°. Hence, the entire core, not just a wedge, was followed. To avoid severe Courant limitations, matter interior to 15 kilometers was followed in 1-D (radial). A 15% decrease in density is a bit larger than yet seen in the calculations of Bazan & Arnett [26], but we imposed no initial aspherical perturbation in velocity, despite the up to Mach 0.25 asymmetries they have seen. (Other calculations we have performed in which the initial asymmetry is solely in velocity, but of a similar magnitude, yielded quantitatively similar effects.) An essential point is that initial asphericities in this simulation grew during collapse, so that the mass column depths in various angular directions diverged. The matter collapsed at different rates in different directions, though pressure forces were transmitted in the angular directions as well that partially smoothed the deviations. The bounce was delayed on the side of the perturbation wedge and the resulting shock bowed out in the wedge direction. The accretion rates through this shock were highly aspherical. To avoid squandering computer cycles in what was merely a "proof-of-principle" calculation, we artificially hardened the emergent neutrino spectrum to facilitate an early explosion. The electron-type neutrino spectra were given an effective degeneracy factor,"$\eta$", of 3, in the upper range of the 1.5–3.0 normally encountered in fits to more realistic spectra [30]. Since neutrino heating drives supernovae, this ignited the explosion within 10 milliseconds of bounce. The subsequent explosion was aspherical not only due to the normal instabilities, but also due to the asphericity of the matter into which the explosion emerged and/or was driven.

Figure 1 depicts the flow late in the explosion. The explosion erupted preferentially through the path of least resistance, *i.e.* in the direction of the wedge that we had imposed.

The wedge collapsed more slowly than the rest of the core. Since neutrino heating drives the explosion, matter heated near the neutrinosphere expands out as if from a reaction chamber. The protoneutron star residue receives a significant impulse à la the rocket effect. In this calculation, the center of the core is held fixed and we infer its recoil speed from the momentum it absorbs. The final core mass is $\sim 1.2$ M$_\odot$ and its recoil speed is depicted in Figure 2. (The quasi-sinusoidal oscillation before bounce at 0.215 seconds seen in Figure 2 is the initial sound wave generated by the perturbation. Had the core not been anchored at the center, the anisotropy at bounce might have been even larger.) Figure 2 shows that the recoil speed grew almost monotonically and reached $\sim 530$ km s$^{-1}$. This is large, but only $\sim 2\%$ of the speed of the supernova ejecta. A small asymmetry in collapse clearly translated into an intrinsic kick, though the initial total momentum was almost zero. Just after bounce, the recoil is in the direction of the wedge, since the rest of the matter bounced first. Afterwards, as the shocked matter starts to squirt through the region of least resistance, the recoil changes sign and grows inexorably to its asymptotic value. The mass motions dominate the recoil. The contribution of neutrino radiation asymmetry to this kick is also depicted in Figure 2 and amounts to a total of $\sim 16\%$. The $\nu_\mu$, $\overline{\nu}_\mu$, $\nu_\tau$, and $\overline{\nu}_\tau$'s contribute 50% and the $\nu_e$'s contribute 30% to the neutrino recoil. The magnitude of the dipole anisotropy of the radiation field varies from 1% to 10% (with an average of $\sim 3\%$ during the first 50 milliseconds) and can change sign. Unfortunately, its behavior is not simple and at the end of the calculation, the dipole is +2%. On balance, it has the same sign as the term due to mass motions, since the neutrino fluxes are larger on the thin, or low-column, side. However, we have yet to demonstrate that this will be true universally.

### III. GRAVITATIONAL RADIATION SIGNATURE

The impulse delivered to the core depends upon the dipole moments of the angular distribution of both the envelope momentum and the neutrino luminosity. The gravitational waveform depends upon the corresponding quadrupole moments. Using the standard quadrupole formula, we derive for the neutrinos the expression,

$$Dh_{zz}^{TT}(t) = \frac{4G}{c^4} \int_{-\infty}^{t} \alpha(t')L_\nu(t')dt',$$

where $D$ is the distance to the supernova, $h_{zz}^{TT}$ is the transverse-traceless and dimensionless metric strain ($z$ is the symmetry axis and the signal's directional dependence has been dropped), $\alpha(t)$ is the instantaneous quadrupole anisotropy, and $L_\nu(t)$ is the total neutrino luminosity [31,32]. Curiously, if $\alpha(t)$ is a constant, $h_{zz}^{TT}$ due to the neutrinos is proportional to the integrated neutrino energy loss! A burst of neutrinos translates into a rapid rise in $h_{zz}^{TT}$. Unfortunately, $\alpha(t)$ in the simulation was not constant, changed sign more than once, and on average was in the opposite direction to the neutrino's dipole term. Furthermore, the net contribution to $h_{zz}^{TT}$ of the neutrinos and the matter were of opposite signs. The magnitude of $\alpha(t)$ averaged $-0.02$ during the first 50 milliseconds after bounce, but achieved $-0.08$ during the early shock break-out burst. The matter's contribution to both the dipole (recoil) and the quadrupole (GW) are of the same sign (as one might expect of an emerging "blob"). Figure 3 depicts the evolution of the strain(s) versus time. Due to the intense and anisotropic early neutrino burst, the neutrino contribution to changes in $h_{zz}^{TT}$ dominates during the first 20 milliseconds. Thereafter, that due to the expanding ejecta asserts itself (though at the end of this calculation, the neutrinos are continuing to contribute). This is true despite the fact that the neutrinos contribute only 16% to the recoil and is a consequence of the relativistic nature of the neutrinos. However, due to the more rapidly changing mass quadrupole around the time of anisotropic bounce, the total energy radiated ($\sim 1.1 \times 10^{-9}$ $M_\odot c^2$) is predominantly due to the mass motions ($\sim 30\%$ is due to the neutrinos).

The gravitational waves are radiated between 10 and 500 Hz. The "memory" in $h_{zz}^{TT}$ seen in Figure 3 is radiated at 10–100 Hz, just at the edge of the LIGO sensitivity range [33]. This memory is a distinctive characteristic of asymmetric collapse and explosion. Using reference [33] and Figure 3, we conclude that the 2'nd-generation LIGO will be able to detect the signal in Figure 3 at a signal-to-noise ratio of 10 from a core collapse anywhere in our galaxy.

A comparison with the results of Mönchmeyer *et al.* [34] for rotating collapse is instructive. In all their models, the asymptotic $h_{zz}^{TT}$ was zero and there was no memory. The maximum energy radiated in gravitational waves was for their Model A and was ∼70 times that for our simulation. However, the maximum $h_{zz}^{TT}$ in their Model A was only ∼3 times that depicted in Figure 3. Interestingly, their Model B experienced a centrifugal bounce at sub-nuclear density, radiated the same total wave energy as our model, but achieved only 70% of its peak $h_{zz}^{TT}$. We see that even without rotation, an asymmetric collapse can result in appreciable gravitational wave emission.

## IV. CONCLUSIONS

The major conclusion of this exploratory, but artificial, simulation is that a small initial core asymmetry can translate after collapse into an appreciable neutron star recoil. Other calculations that we have performed imply that this asymmetry/recoil correlation is generic. However, whether such asymmetries are themselves generic has yet to be demonstrated. This recoil is correlated with an appreciable gravitational wave signal with "memory," and both are correlated in a complicated fashion with the neutrino emissions. The simultaneous detection of these signatures would provide direct and unique information concerning the dynamics of the supernova mechanism. Since accretion-induced collapse (AIC) is not preceded by the convective burning stages characteristic of the final hours of the core of a massive star, the initial asymmetries in the two contexts may be quite different. Consequently, the proper motions of AIC neutron stars and those of neutron stars from massive stars may be systematically different, with those of the latter being on average higher. This may be of relevance in the study of globular cluster neutron star binaries.

The simulation highlighted in this paper is but an idealized study of the character of anisotropic collapse and explosion. It is in no way definitive. To explore these phenomena more credibly will require a realistically aspherical initial core, 3-D simulations, and much better neutrino transfer. Nevertheless, we have reproduced the correct magnitude of the observed pulsar proper motions and have discovered potentially rich correlations between recoils, gravitational radiation, and neutrino emissions. Given the stochastic nature of the processes we highlight here, we expect that Nature provides not a single high kick speed, but a broad distribution of speeds. These will depend upon the degree and character of the initial asymmetries, the initial rotation structure, the duration of the delay to explosion, the progenitor density profiles, and chance.

## ACKNOWLEDGMENTS

The authors would like to thank Grant Bazan, Willy Benz, Sam Finn, Icko Iben, Thomas Janka, Dan Kennefick, and Kip Thorne for stimulating conversations during the germination of this work and the N.S.F. for financial support under grant # AST92-17322. They would also like to thank the Pittsburgh Supercomputer Center for making available to them its C90, without which these calculations could not have been so expeditiously completed.

FIGURES

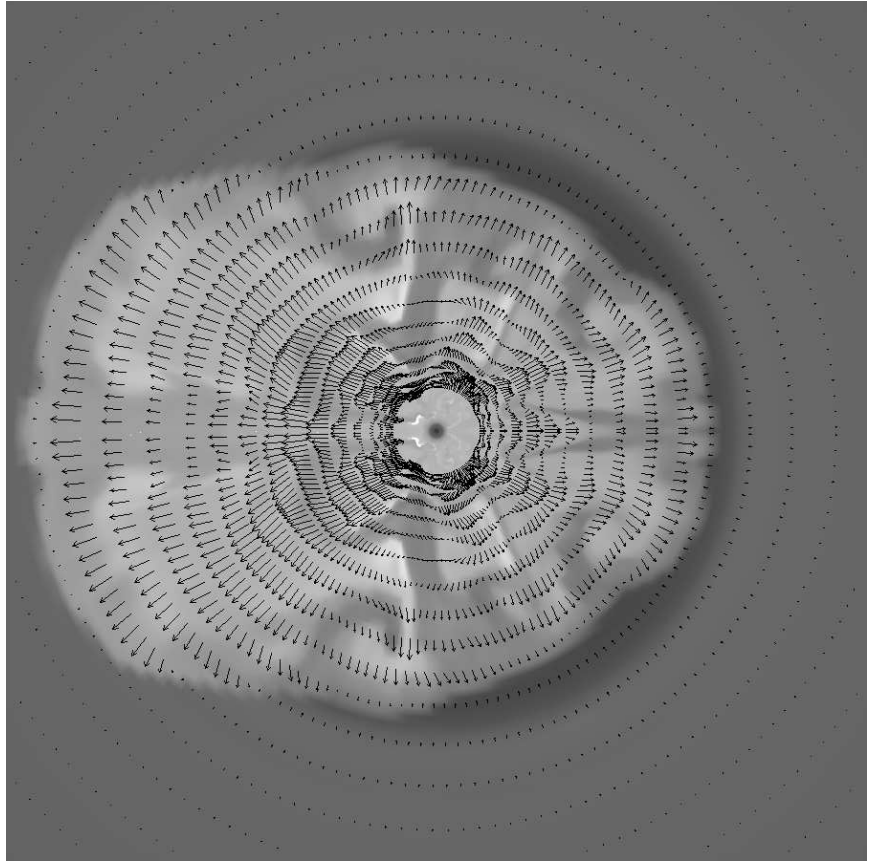

FIG. 1. A grey-scale rendering of the entropy distribution at the end of the simulation, about 50 milliseconds into the explosion. Note the pronounced pole–to–pole asymmetry in the ejecta and the velocity field (as depicted with the velocity vectors). The physical scale is 2000 km from the center to the edge. Darker color indicates lower entropy and $\theta = 0$ on the bulge side of the symmetry axis.

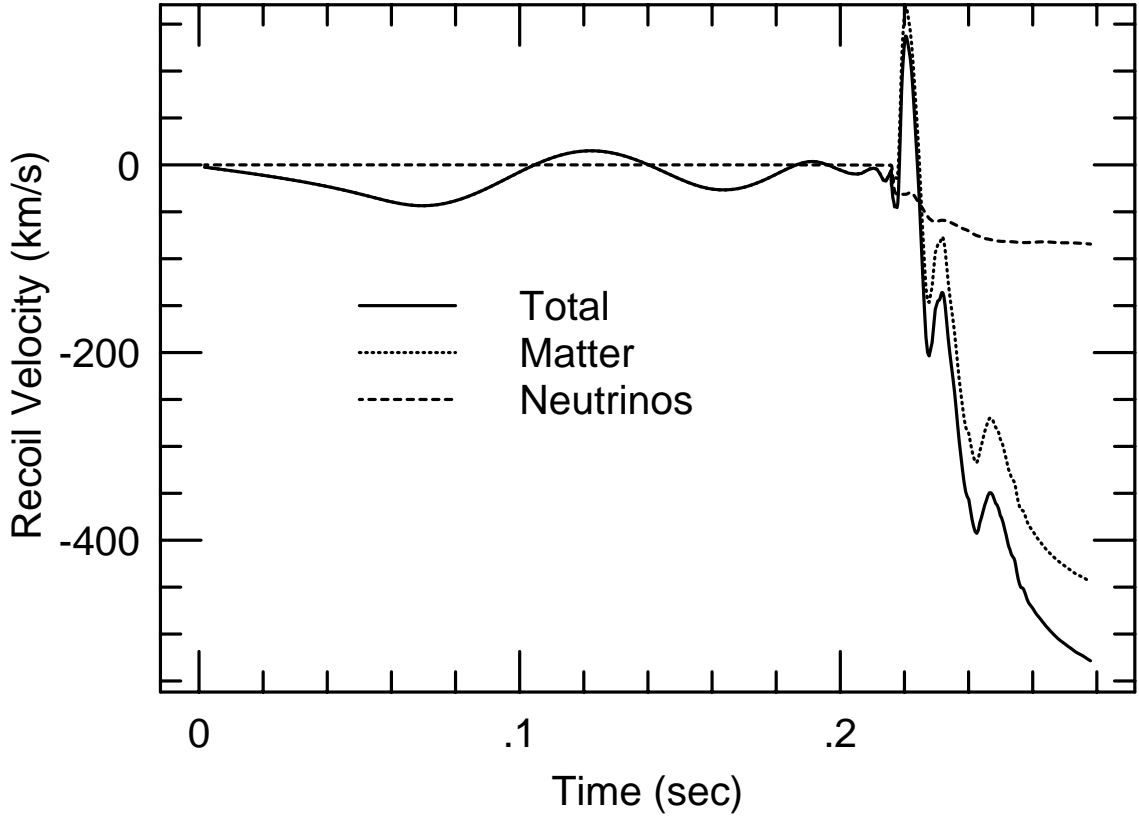

FIG. 2. The inferred recoil speed (in km s$^{-1}$) imparted to the core versus time (in seconds) for the simulation highlighted in this paper. The initial momentum is approximately zero, but grows systematically after bounce in the direction opposite to the artificial wedge, cut into the core to mimic an asymmetry just before collapse. Shown are the total recoil (solid) and the contributions due to the neutrino emission anisotropy (dashed) and the ejecta motions (dotted).

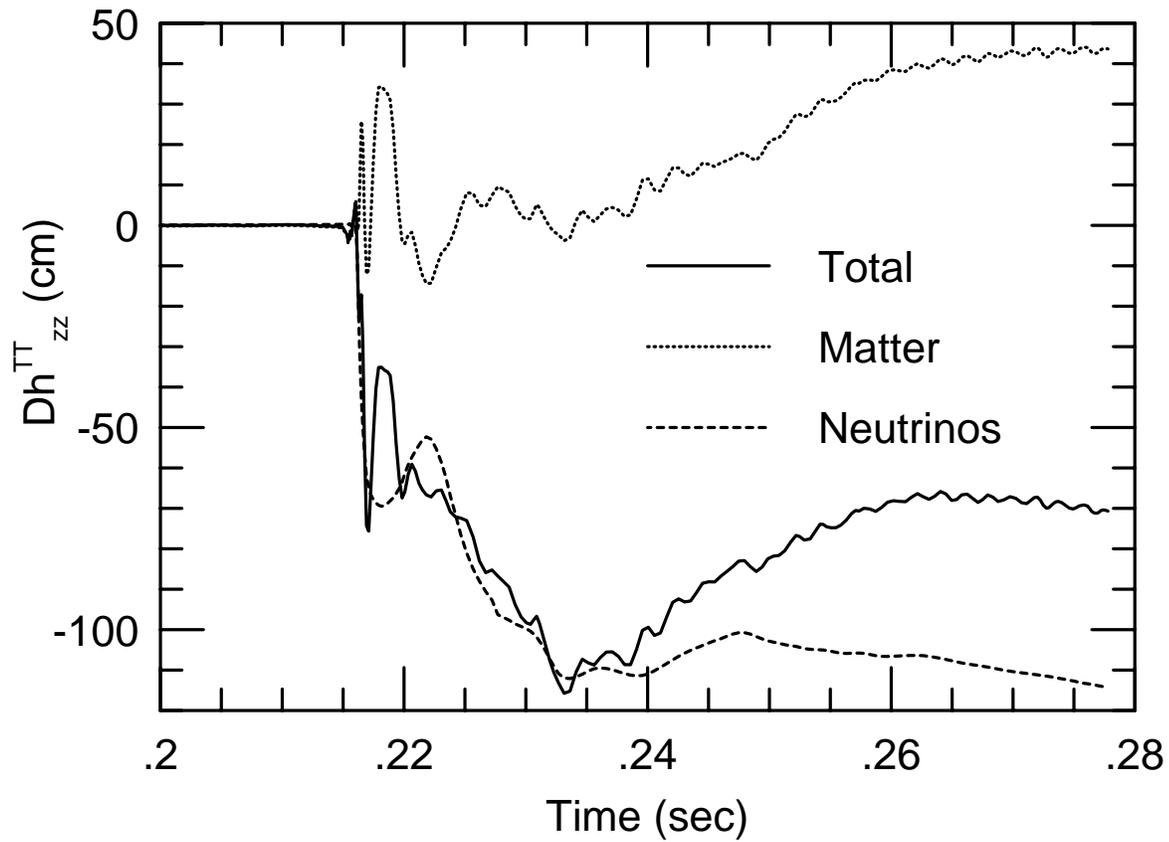

FIG. 3. The gravitational wave strain, $h_{zz}^{TT}$, times the distance to the supernova, D, versus time (in seconds). Core bounce is at 0.215 seconds. The total, matter, and neutrino waveforms are rendered with the solid, dotted, and dashed lines, respectively.